\documentclass[twocolumn,showpacs,preprintnumbers,amsmath,amssymb]{revtex4}
\usepackage{graphicx}% Include figure files
\usepackage{dcolumn}% Align table columns on decimal point
\usepackage{bm}% bold math

\begin{document}

\title{Denaturation Patterns in Heterogeneous DNA}

\author{ Marco Zoli }
\affiliation{
School of Science and Technology - CNISM \\  Universit\`{a} di Camerino, I-62032 Camerino, Italy \\ marco.zoli@unicam.it}

\date{\today}

\begin{abstract}
The thermodynamical properties of heterogeneous DNA sequences are computed by path integral techniques applied to a nonlinear model Hamiltonian. The base pairs relative displacements are interpreted as time dependent paths whose amplitudes are consistent with the model potential for the hydrogen bonds between complementary strands. The portion of configuration space contributing to the partition function is determined, at any temperature, by selecting the ensemble of paths which fulfill the second law of thermodynamics. For a short DNA fragment, the denaturation is signaled by a succession of peaks in the specific heat plots while the entropy grows continuously versus $T$. Thus, the opening of the double strand with bubble formation appears as a smooth crossover due to base pair fluctuation effects which are accounted for by the path integral method.
The multistep transition is driven by the AT-rich regions of the DNA fragment. The base pairs path ensemble shows an enhanced degree of cooperativity at about the same temperatures for which the specific heat peaks occur. These findings establish a link between microscopic and macroscopic signatures of the transition. The fractions of mean base pair stretchings are computed by varying the AT base pairs content and taking some threshold values for the occurrence of the molecule denaturation.
\end{abstract}


\maketitle

\section*{I. Introduction}

Partial separation of the DNA double helix is fundamental in many processes relevant for biological functioning such as transcription and replication of the genetic information \cite{wart}. Also the packing of long DNA strands into nucleosomes seems related to the local opening of a double helix segment which may provide the key mechanism for loop formation  \cite{marko}. Gene transcription is possible as the hydrogen bonds, linking the pair bases on the two complementary strands, can break and expose the bases for chemical reaction. The region of open base pairs ({\it bps}), the transcription bubble, is generally localized and characterized by large amplitude fluctuations known as the {\it breathing} of DNA. These  observations have suggested that the {\it bps} hydrogen bonds are intrinsically nonlinear \cite{proh} thus putting some constraints on the modeling of the double helix dynamics and strands separation, the DNA denaturation. The latter is driven experimentally either by increasing the temperature or reducing the proton concentration in the solvent so that the repulsion between negative phosphate groups on the two strands is less screened. Also adsorption of DNA on a surface affects the denaturation properties, a process widely used in biotechnologies \cite{felg,raed}. Thermally induced bubbles can be several {\it bps} long even at about room temperature and extend by increasing $T$, leading to the DNA melting once the complete strand separation occurs. Such process is made evident by a sharp increase in the UV absorbance \cite{inman} of the DNA solution due to the reduction of both base pairing and stacking (along the strand) upon denaturation. In fact substantial differences occur in the UV absorption profiles for synthetical {\it homogeneous} and natural {\it heterogeneous} DNA: while the former denaturates within a narrow temperature range, the latter shows multiple steps transitions \cite{blake} according to patterns which depend both on the length and on the sequence \cite{carlon,zocchi,krueg,joy3}, that is on fraction and specific order of the strongly bonded {\it guanine-cytosine} and weakly bonded {\it adenine-thymine} base pairs. However, as a common signature to all different DNA structures, denaturation is a highly cooperative phenomenon involving a sizeable number of {\it bps}. This follows from the fact that the thermal disruption of a specific inter-strand hydrogen bond decreases the overlap between $\pi$ electron orbitals of the organic rings in the bases and favors the unstacking of intra-strand adjacent bases which, in turn, breaks the next hydrogen bond and ultimately opens a bubble in the double helix \cite{yakov}. The role of cooperativity effects in DNA has been recognized since long \cite{poland,azbel1,fisher,richard} and introduced phenomenologically in Ising-like two state models in which the {\it bps} are either closed or open. Such models have been applied to represent melting transitions occuring step by step in heterogeneous DNA fragments \cite{wada,azbel2}.  Later on Hamiltonian models, in which the potential energy is continuous function of the distance between the bases \cite{pey1}, have proposed a microscopic origin for cooperativity by relating it to the anharmonic character of the intra-strand stacking potential \cite{pey2}. The latter has been found responsible for a denaturation with an entropy jump corresponding to an effective latent heat reminiscent of a first order phase transition in homogeneous DNA \cite{theo}. However, no consensus has been reached so far regarding the nature of the transition, whether first or second order \cite{peliti,causo,stella,hanke1,bar,rud,manghi,hanke,santos}.

The Peyrard-Bishop-Dauxois (PBD) anharmonic model \cite{pey2} has also proved to be consistent with a multistep melting envisaged by experiments in heterogeneous DNA \cite{cule} and with the formation of temporary, sequence dependent openings observed by S1 nuclease cleavage experiments \cite{choi}. Instead, some discrepancies have been pointed out between the PBD predictions and the denaturation curves of specific heterogeneous sequences \cite{pey3,zocchi1,zocchi2} indicating that improvements in the theoretical modeling are still necessary.
Due to the huge number of degrees of freedom, fully atomistic representations for sizeable segments of DNA require prohibitive computational time. Accordingly, several mesoscopic models have been developed to account for the essential interactions which determine structural stability, dynamics and denaturation of the molecule \cite{druk,joy1,sal,kno,samb}.

In a recent work \cite{io}, the imaginary time path integral formalism has been applied to the PBD Hamiltonian to investigate the occurrence of thermal denaturation in homogeneous DNA. The transverse stretchings of the {\it bps} with respect to the ground state have been treated as one-dimensional paths $x(\tau)$ depending on the imaginary time $\tau$ whose range is set by the inverse temperature \cite{fehi}. A path is defined by a set of Fourier coefficients and a single base pair displacement is taken at a specific $\tau_i$. Then, an ensemble $\{x(\tau_i)\}$ $\,( i=\,1, N )\,$ represents a configuration for the DNA molecule made of N {\it bps} and, by varying the Fourier coefficients, one builds all the possible molecule states at a selected temperature. While {\it in principle} the path integral is obtained by summing over all DNA configurations, the model potential poses lower and upper bounds on the specific bps elongations which naturally restrict the path phase space for the computation of the partition function. The method accounts for the highly cooperative character of the denaturation which appears as a smooth second order transition in homopolymer DNA.

In this paper the path integral formalism \cite{feyn} is extended to heterogeneous DNA and, in particular, to short fragments which  are \emph{both} technologically interesting for fabrication of DNA chips \cite{fiche} \emph{and} theoretically relevant due to the enhanced role of fluctuations far from the thermodynamic limit (finite and small $N$) \cite{manghi1,bonnet}. Due to the direct integration over the {\it bps} degrees of freedom, the path integral method naturally incorporates fluctuation effects and seems therefore particularly promising in dealing with finite size DNA fragments.
The PBD Hamiltonian and the generalities of the path integral approach are presented in Section II. The thermodynamical properties for some specific DNA sequences are discussed in Section III together with the computation of the fractions of open {\it bps} versus temperature. Some final remarks are made in Section IV.

\section*{II. Theory}

\subsection*{A. Hamiltonian Model}

The PBD Hamiltonian, originally introduced for homogeneous DNA \cite{pey2}, is usually extended to represent a chain of N heterogeneous {\it bps} as follows

\begin{eqnarray}
& & H =\, \sum_{n=1}^N \biggl[ {{\mu \dot{y}_{n}^2} \over {2}} +  V_S(y_n, y_{n-1}) + V_M(y_n) \biggr] \, \nonumber
\\
& & V_S(y_n, y_{n-1})=\, {K \over 2} g(y_n, y_{n-1})(y_n - y_{n-1})^2 \, \nonumber
\\
& & g(y_n, y_{n-1})=\,1 + \rho \exp\bigl[-\alpha(y_n + y_{n-1})\bigr]\, \nonumber
\\
& & V_M(y_n) =\, D_n \bigl(\exp(-a_n y_n) - 1 \bigr)^2 \, ,
\label{eq:1}
\end{eqnarray}

where $y_n$ is the transverse stretching at the {\it n-th} site and measures the relative pair mates separation from the ground state position.
The model is essentially one-dimensional as the longitudinal displacements, being much smaller than the transverse stretchings, are not taken into account \cite{wart}.
The boundary condition $y_0=\,y_N$  closes the chain into a loop whereas, in the case of an open end chain with $N+1$ {\it bps}, the single particle energy for $y_0$ should be added to Eq.~(\ref{eq:1}).  $\mu$ is the reduced mass of the bases which is assumed identical both for GC and AT {\it bps}. This is a relevant limitation of the model \cite{blake,yakov} which is mirrored also in the stacking potential $V_S(y_n, y_{n-1})$ whose parameters $K$ and $\rho$ are independent of the type of bases at $n$ and $n-1$. In fact, $K=\, \mu \nu^2$ with $\nu$ being the harmonic phonon frequency.

$\rho$ ($ > 0$) accounts for the anharmonicity in the stacking of nearest neighbors {\it bps}. When the molecule is closed, $y_n \,, y_{n-1} \ll \alpha^{-1}$, the effective stacking coupling is $K(1 + \rho)$.
Whenever either $y_n > \alpha^{-1}$ or $y_{n-1} > \alpha^{-1}$, the corresponding hydrogen bond breaks and the electronic distribution around the two pair mates is modified. Accordingly in Eq.~(\ref{eq:1}), $g(y_n, y_{n-1}) \sim 1$ and the effective stacking coupling (along each strand) between neighboring bases drops to $K$.
Then, also the adjacent base tends to open as both bases are less closely packed along their respective strands. This is the microscopic origin of the cooperative character (emphasized in the Introduction) of the interactions which determine the formation of a region with open {\it bps}. The interplay between anharmonicity and cooperativity  is thus at the heart of the PBD model through the form of the stacking potential.

However the form for $V_S(y_n, y_{n-1})$ in Eq.~(\ref{eq:1}) is not unique and other potentials have been proposed which also account for the finiteness of the stacking energy at large intra-strand base separations \cite{joy1}. Typical values for DNA models with intermediate anharmonicity are taken hereafter, $K=\,60 meV {\AA}^{-2}$, $\alpha=\,0.35 {\AA}^{-1}$ and $\rho=\,1$ \cite{theo,joy2}. As the parameters are site independent, it follows that the present discussion is neglecting stacking hetereogeneities \cite{blake}. The latter may slightly affect the melting temperatures of specific portions of the chain \cite{joy3} although they are not expected to modify the nature of the denaturation crossover. The quantitative effects of the stacking hetereogeneities are left for next investigations.

\begin{figure}
\includegraphics[height=7.0cm,angle=0]{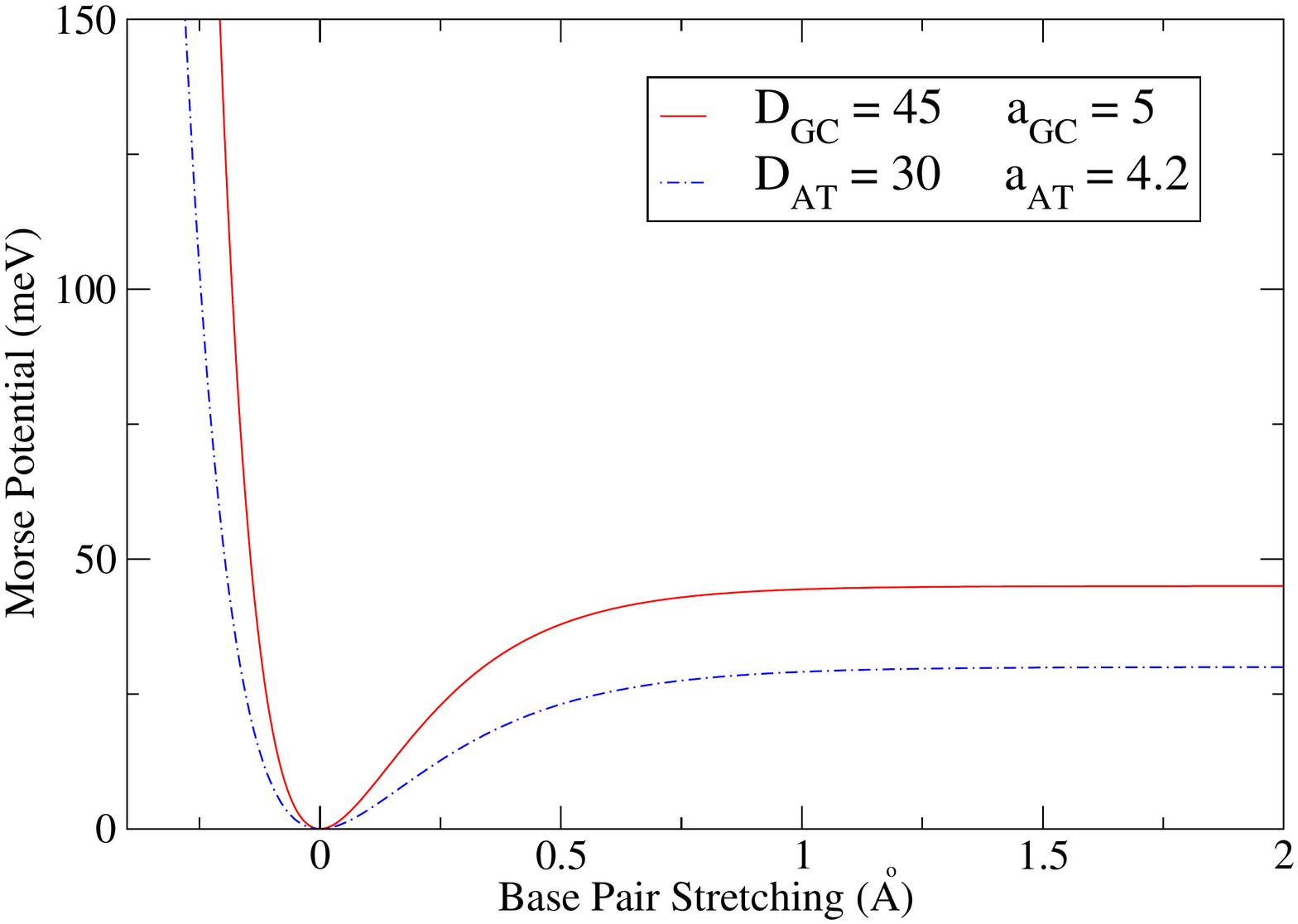}
\caption{\label{fig:1}(Color online) Morse potential $V_M$ versus base pair relative separation.  The potential parameters $D_n$ (in $meV$) and $a_n$ (in ${\AA}^{-1}$) are taken for both GC and AT base pairs.}
\end{figure}

Instead, heterogeneity is present in the Morse potential $V_M(y_n)$ which models the hydrogen bond link between bases on complementary strands \cite{proh1,proh2}. Depth $D_n$ and width $a_n$ of the potential differ for weakly bonded AT {\it bps} and strongly bonded GC {\it bps}. Fig.~\ref{fig:1} shows $V_M(y_n)$ for the parameters used in the following calculations. While hydrogen bonds may vary in a considerable range \cite{parri}, those in DNA are typically described by taking energies per bond of $\sim 15 - 25 meV$ \cite{pey4}. I assume here the lower bound taking $D_{AT}=\,30 meV$ and $D_{GC}=\,45 meV$ thus accounting for the fact that AT and GC {\it bps} have two and three bonds respectively.  $a_n$ sets the spatial cutoff beyond which the {\it bps} tend to open. The values $a_{AT}=\,4.2 {\AA}^{-1}$ and $a_{GC}=\,5 {\AA}^{-1}$, ensure that transverse stretchings are somewhat stiffer for GC than for AT {\it bps} although even larger values for $a_{GC}$ are found in the literature \cite{campa}.

In spite of some arbitrariness in the parameters choice, the shape of $V_M(y_n)$ captures the fundamental features of the many body interactions at play between the opposite strands. The repulsion of the negatively charged phosphate groups is described by the hard core that the base pair mates experience by coming too close to each other ($y_n < 0$). On the opposite side, when the relative separation grows above a given {\it threshold}, the pair opens and the force between the mates vanishes consistently with the plateau at the dissociation energy encountered for large $y_n$. However precisely the plateau, at about $y_n > 1{\AA}$ in Fig.~\ref{fig:1}, reveals a drawback of the model: when all {\it bps} in the chain open, the two strands separation can grow in principle to infinite with no further effort as the potential energy is flat \cite{note}. Thus, the PBD Hamiltonian assumes a single chain in a infinite solution whereas experiments deal with DNA in a solvent structure at finite concentration hence, recombination of separated strands in solution is possible.
Here is a case of biomolecule whose structure depends on the strong interaction with the environment, a challenge to theoretical investigation \cite{allah}. Specifically, reconciling model to experiment requires some restrictions of the configuration space which have been attempted either by methods based on molecular dynamics \cite{pey5} and Monte-Carlo simulations \cite{ares} or by truncating the kernel domain in the transfer integral method \cite{scala} to prevent the two strands from going infinitely apart \cite{zhang}.
In the path integral formalism, proposed in Ref. \cite{io} and briefly outlined in the next Subsection, the confinement of the phase space for the $\{y_n\}$ is naturally incorporated in the computation after imposing a macroscopic constraint to the evolution of the system which is driven by the temperature.

It is also worth pointing out that the lower bound confinement for the $\{y_n\}$, physically due to the hard core potential, ensures that the {\it bps} paths are self-avoiding at complementary sites along the strands. In fact base pair mates do not overlap.  However we recognize that the system in Eq.~(\ref{eq:1}), lacking of the rotational degrees of freedom, does not capture the helicoidal structure of the molecule which should be realistically embedded in the three dimensional space. Accordingly, also self-avoidance is only partially considered in our investigation as excluded volume effects due to interactions between bubbles and bounded segments in three dimensions are not taken into account. This effect has been shown to be relevant in polymer network theories to drive a sharp denaturation transition at least in homogeneous DNA \cite{causo,stella}.  On the other hand, the path integral approach to the Hamiltonian in Eq.~(\ref{eq:1}) accounts for  all {\it bps} fluctuations at any $T$ and permits to include in the computation the two competing tendencies of the system: the energetic gain associated to the (bounded) double strands configuration and the entropic gain due to the large number of configurations available for open strands.

\subsection*{B. Path Integral Method}

The idea underlying the path integral method is that of mapping the real space model in Eq.~(\ref{eq:1}) onto the imaginary time scale.  Accordingly, the transverse stretching $y_n$ is represented by a one dimensional path $x(\tau_i)$ with $\tau_i \in [0, \beta]$ and $\beta$ being the inverse temperature:

\begin{eqnarray}
& &y_n \rightarrow x(\tau_i), \, \, \, y_{n-1} \rightarrow x(\tau')\, , \, \tau' = \tau_i - \Delta \tau \, , \, \nonumber
\\
& &n =\, 0\,,N \,;\, i =\,1\,,N_\tau + 1 \, \, . \,
\label{eq:3}
\end{eqnarray}

Thus, at any given temperature, the finite size system of $N + 1$ {\it bps}, is described by $N_\tau + 1$ $(N_\tau \equiv N )$ paths each of them taken at a specific $\tau_i$ along the time axis.
Along the DNA strands only adjacent {\it bps} stacking interactions are considered. Accordingly, $\tau_i$ and $\tau'$ in Eq.~(\ref{eq:3}), are first neighbors separated by $\Delta \tau $ in the discrete imaginary time lattice.

I am assuming periodic boundary conditions, $x(0)=\,x(\beta)$, for all paths analogously to those imposed for the 1D finite chain described by Eq.~(\ref{eq:1}) \cite{pey2}.
Then, periodicity ensures that a molecule configuration is given by $N_\tau$ paths and the retardation is: $\Delta \tau =\,\beta / N_\tau$. Further, any path $x(\tau_i)$ can be  expanded in Fourier series with cutoff $M_F$

\begin{eqnarray}
x(\tau_i)=\, x_0 + \sum_{m=1}^{M_F}\Bigl[a_m \cos(\omega_m \tau_i) + b_m \sin(\omega_m \tau_i) \Bigr] \, , \,
\label{eq:3a}
\end{eqnarray}

with $\omega_m =\, {{2 m \pi} / {\beta}}$.
Using Eq.~(\ref{eq:3a}) has an important physical interpretation: for any choice of coefficients $\{x_0 , a_m , b_m\}$, a single configuration $\{x(\tau_i)\}$ for the DNA fragment is built at a given temperature. As such coefficients can be varied in the phase space, many different configurations are possible at the same temperature  each of them being a copy of the molecule in the ensemble.
Thus, integration over the path coefficients amounts to sample the molecule configuration space and, in turn, to  account for the possible evolutions of the N {\it bps} system in going between the time points $0$ and $\beta$.

As the trajectories are closed paths, the path integral yields the imaginary time partition function \cite{fehi} which is given by

\begin{eqnarray}
& &Z=\,\oint \mathfrak{D}x\exp\bigl[- A\{x\}\bigr]\, \nonumber
\\
& &A\{x\}=\,\int_0^\beta d\tau \Bigl[{\mu \over 2}\dot{x}(\tau)^2 + V_S(x(\tau),x(\tau')) + V_M(x(\tau)) \Bigr] \, \nonumber
\\
& &\oint \mathfrak{D}x\equiv {1 \over {\sqrt{2}\lambda_\mu}}\int dx_0 \prod_{m=1}^{M_F}\Bigl({{m \pi} \over {\lambda_\mu}}\Bigr)^2 \int da_m \int db_m \, \, , \, \nonumber
\\
\label{eq:3b}
\end{eqnarray}

where $A\{x\}$ is the Euclidean action for the molecule in Eq.~(\ref{eq:1}) after applying the mapping in Eq.~(\ref{eq:3}). The molecule state $\{x\}$ corresponds to a specific set of Fourier coefficients. In practice, the $d\tau$ integral is replaced by  $\sum_{i=\,1}^{N_\tau}$ (and $x(\tau) \rightarrow x(\tau_i)$) which has to be sufficiently dense to make the action numerically stable. This poses a constraint to the application of the method to very short DNA fragments.
Hereafter I take $N_\tau =\,100$ while a possible extension of the method to molecules of arbitrary length will be mentioned in the Conclusion.

$\mathfrak{D}x$ is the measure of integration which normalizes the free particle action

\begin{eqnarray}
\oint \mathfrak{D}x\exp\Bigl[- \int_0^\beta d\tau {\mu \over 2}\dot{x}(\tau)^2  \Bigr] =\,1
\label{eq:6} \,
\end{eqnarray}

and $\oint$ denotes integration over closed particle trajectories \cite{kleinert}.
$\lambda_\mu$ is the thermal wavelength whose form in general depends on the model whether quantum or classical. The latter is appropriate to the occurrence of DNA denaturation. Then, the time derivative $\dot{y}_n$ (Eq.~(\ref{eq:1})) maps  onto the imaginary time derivative $\dot{x}(\tau)$ (Eq.~(\ref{eq:3b})),  the proper replacement being:  $d/dt \rightarrow ({\nu \beta}) d/d\tau$ hence, ${\lambda_\mu}=\,\sqrt{{\pi } / {\beta K}}$.

The above mentioned truncation of the configuration space is intrinsic to the path integral method as the computation of Eq.~(\ref{eq:3b}) requires a cutoff in the Fourier coefficients integration \cite{io2,io3}. The latter has to be consistent with the physics contained in the model potential. Paths $x(\tau_i) \sim 0$ represent the equilibrium configuration for the double helix corresponding to the minimum $V_M(x(\tau_i))$. Then, {\it qualitatively}, one may argue that too large coefficients would produce: {\it i)} too negative path amplitudes in Eq.~(\ref{eq:3a}) which are forbidden by the electrostatic repulsion between the sugar-phosphate backbones \cite{xu}; {\it ii)} too positive paths which are anyway unphysical as the two strands separation has an upper bound. As Fig.~\ref{fig:1} makes clear, paths associated to AT {\it bps} can sample a spatial range somewhat broader than paths describing GC {\it bps} whose bonds are stiffer. Incorporating all these requirements it is found that the suitable set of paths should be searched among the $x(\tau_i) \in [x_{min}, x_{max}]$, with $x_{min} \sim -0.2{\AA}$ and $x_{max} \sim 6{\AA}$ for the temperature window hosting denaturation effects. More negative paths would make a vanishing contribution to the partition function (making the free energy $F$ of the system numerically unstable) while larger positive paths would not affect the free energy derivatives. After setting the framework, the {\it quantitative} determination of the paths configuration space is carried out by imposing the fulfillment of the second law of thermodynamics.

The free energy derivatives, presented in the next Section, are obtained by $F=\, -{\beta^{-1}}\ln Z$ with $Z$ given in Eq.~(\ref{eq:3b}).

Thus Eq.~(\ref{eq:3b}) is computed, at an initial temperature $T_I$, for a given path ensemble defined (at any $\tau_i$) by the number of integration points over the Fourier coefficients. The path ensemble is temperature dependent. Then, at any larger $T$, the numerical code re-determines the contribution to Z and calculates the free energy derivatives.
If, for a given number of integration points, the growing entropy constraint is not fulfilled then the size of the path ensemble is increased. The procedure is reiterated until a {\it minimum number of paths} is found such that the entropy grows versus $T$.  This method sets the $T$-dependent size of the ensemble, $N_{eff}$, whose paths satisfy boundary conditions and macroscopic physical constraints. These are the good paths included in the computation.  $N_{eff}$ is the number of different trajectories followed by a single base pair stretching in the configuration space. As the procedure holds for any $\tau_i$, the total number of paths contributing to the thermodynamics is $N_\tau \times N_{eff}$ whose value sets the overall system size. Good numerical convergence has been found taking $N_{eff} \sim 47000$ at $T_I=\,260K$ \cite{io} and no significant effect arises by further increasing the initial size of the path ensemble.

\section*{III. Denaturation Curves}

In heterogeneous DNA,  AT-rich portions of the molecule tend to open at lower temperatures than GC-rich regions. However openings occurring at lower temperatures extend also well inside the GC domains indicating a role for nonlocal effects in shaping multistep denaturation patterns \cite{angelov}. The sequence pattern is particularly relevant in relatively short segments made of a few tens of  {\it bps} which is the relevant scale for those transcription starting domains where the genes are read. As transcription and other biological phenomena require formation of open domains, theoretical modeling faces the questions to define when:  {\it a)} a base pair is open, {\it b)} a molecule is open. Here I consider the statistical average of the $i-th$ base pair elongation as given by

\begin{eqnarray}
& &< x(\tau_i) >=\,Z^{-1}\oint \mathfrak{D}x x(\tau_i) \exp\bigl[- A\{x\}\bigr] \,. \,
\label{eq:7}
\end{eqnarray}

Eq.~(\ref{eq:7}) is computed by summing over those {\it good paths} in the configuration space which fulfill the growing entropy constraint as described in the previous Section. Then a base pair is open if: $<x(\tau_i)>\, \geq\, \zeta$, where the {\it threshold} $\zeta$ is an arbitrary parameter at this stage. Further, the fraction of open {\it bps} is defined as

\begin{eqnarray}
& &f =\, {1 \over {N_\tau}}\sum_{i=1}^{N_\tau} \theta\bigl(< x(\tau_i) > - \zeta \bigr) \, , \,
\label{eq:8}
\end{eqnarray}

where $\theta(\bullet)$ is the Heaviside step function.
Accordingly, the size of the local openings is measured by $f > 0$ while a molecule is entirely open if  $\,f=\,1$.
This does not imply that all molecule configurations in the ensemble are denaturated: I am assuming that a DNA molecule may exist in many different configurations which have to be Boltzmann weighted to get the ensemble average of physically relevant quantities. If all the averaged elongations exceed a given threshold then the two strands separate.

Somewhat different definitions for $f$ appear in the literature \cite{pey3,ares} with some authors arguing that the UV absorption signal does not relate to the mean {\it bps} stretching hence, the sum in Eq.~(\ref{eq:8}) should be made over the statistical averages  $< \theta\bigl(x(\tau_i)  - \zeta \bigr) >$.
While this point should be investigated in connection with available experiments for ensembles of short molecules \cite{zocchi1}, here the focus is rather on the trend of the path integral model predictions for a single molecule.

\begin{figure}
\includegraphics[height=7.0cm,angle=0]{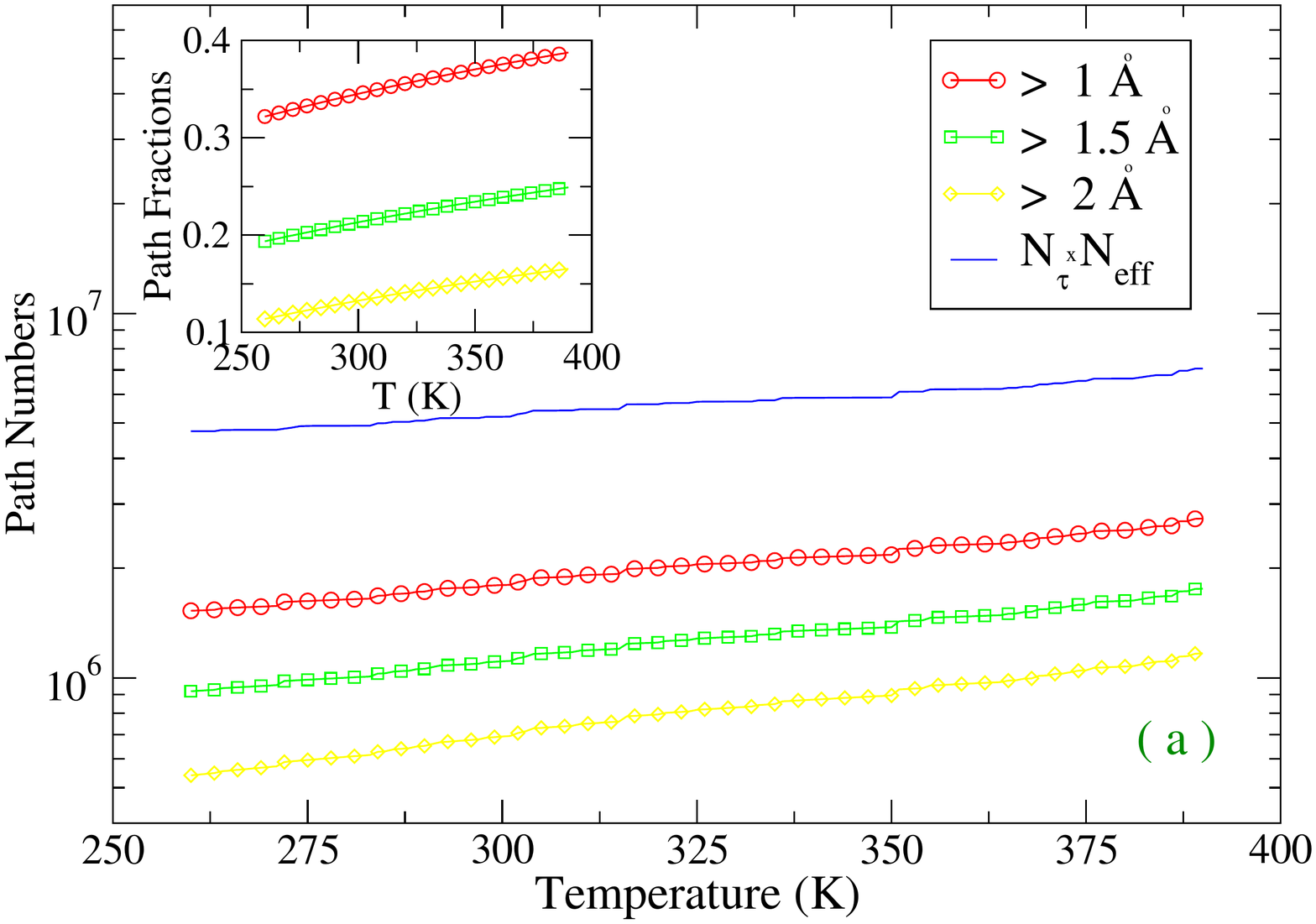}
\includegraphics[height=7.0cm,angle=0]{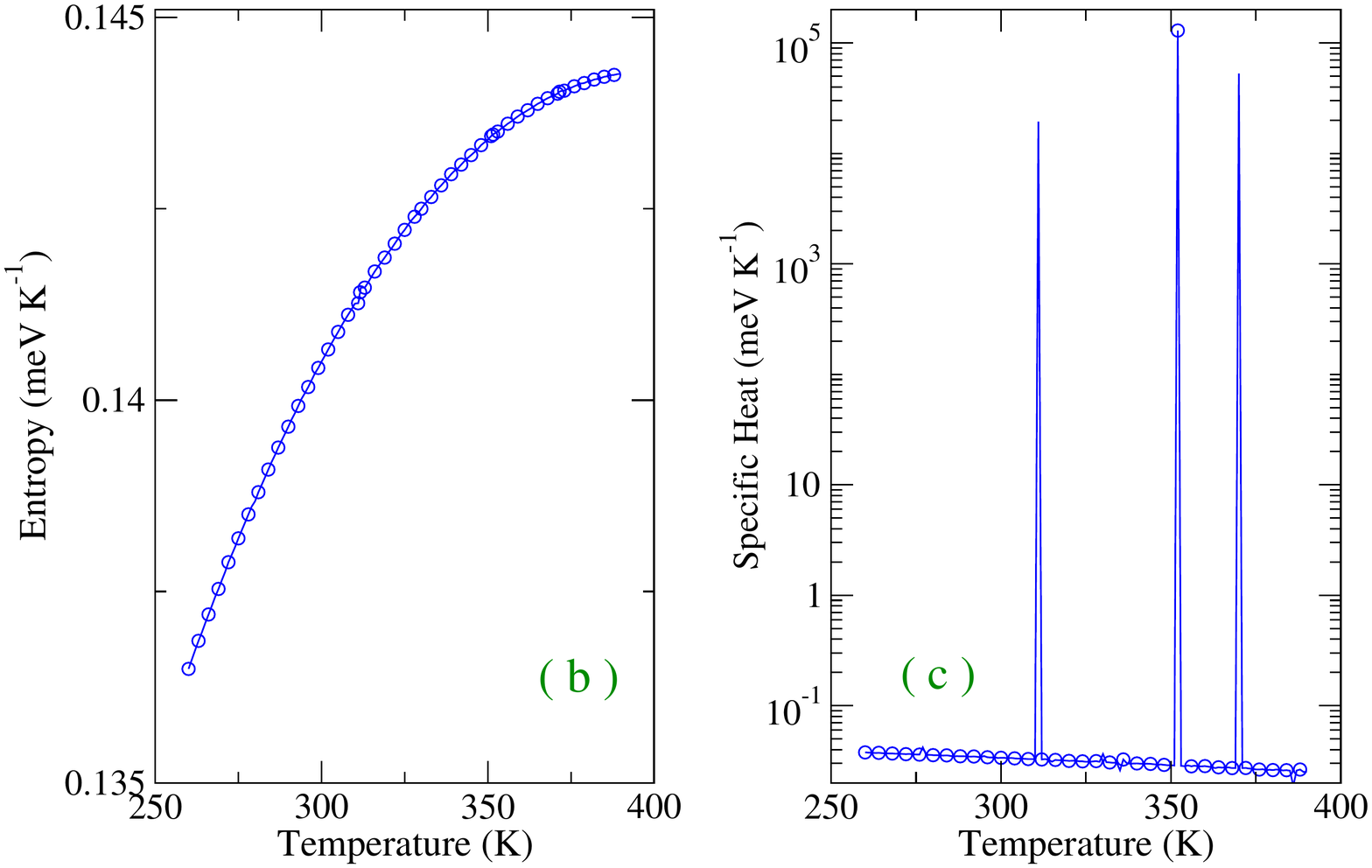}
\includegraphics[height=7.0cm,angle=0]{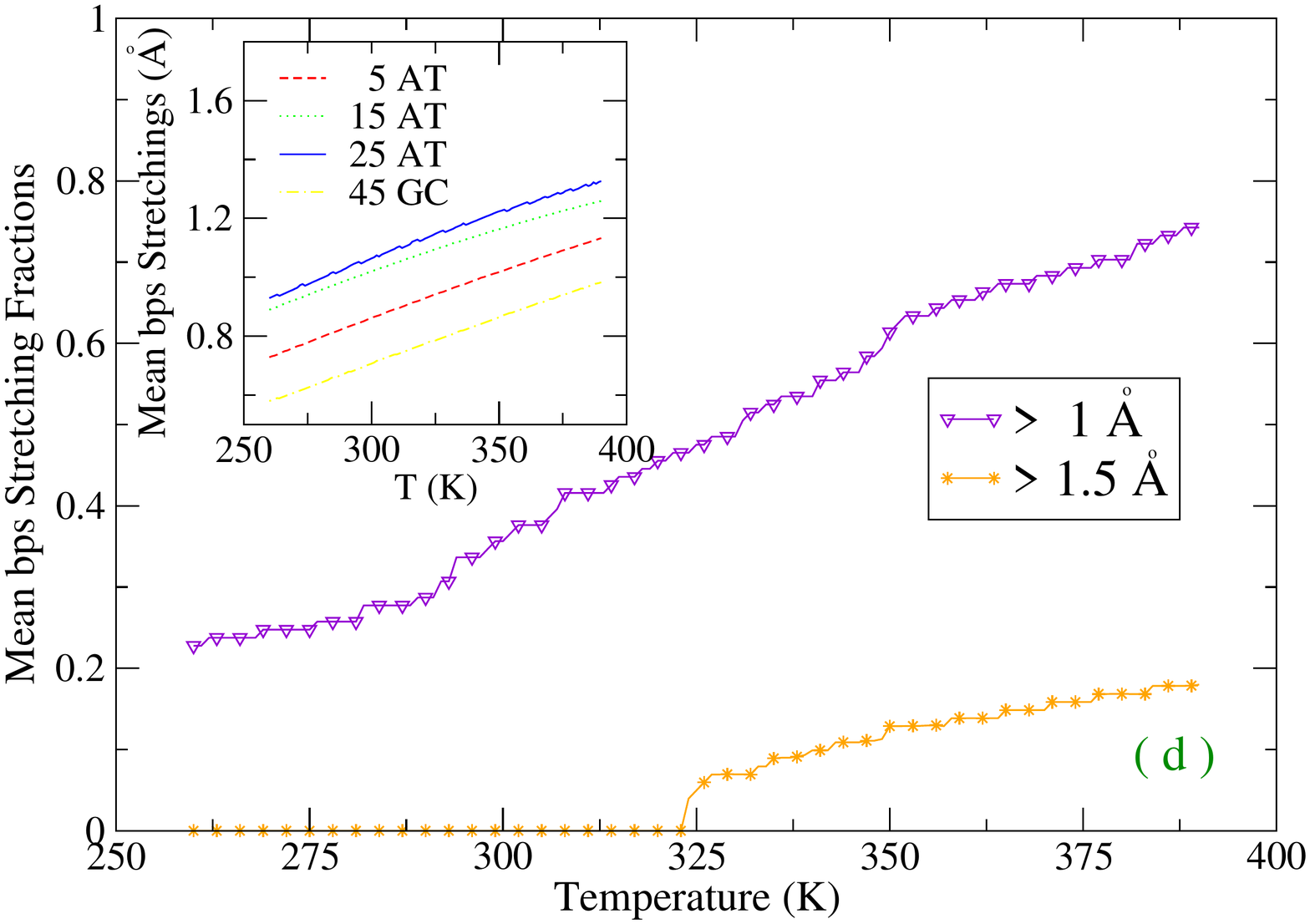}
\caption{\label{fig:2}(Color online) Sequence L48(AT30)+GC[49-100] in the temperature range which shows denaturation. (a) Number of paths (for a single base pair stretching) larger than $1{\AA}$, $1.5{\AA}$ and $2{\AA}$;  Total Number of paths ($N_\tau \times N_{eff}$) contributing to the partition function. Inset: number of paths (per base pair) whose amplitude is larger than $1{\AA}$, $1.5{\AA}$ and $2{\AA}$ respectively, over $N_\tau \times N_{eff}$.  (b) Entropy versus temperature.  (c) Specific Heat versus temperature. (d) Fractions of mean base pair stretchings calculated by Eq.~(\ref{eq:8}) for $\zeta=\, 1{\AA}$ and $\zeta=\,1.5{\AA}$ respectively. Inset: mean base pair stretchings at four specific sites.}
\end{figure}

First I consider a GC-rich molecule with 100 {\it bps} whose sequence is:

\begin{eqnarray}
GC &+& 6AT + GC + 22AT + 4GC + AT + 4GC  \nonumber
\\
&+& AT + 8GC + [49-100]GC
\label{eq:9}
\end{eqnarray}

The index $i$ (Eqs.~(\ref{eq:7}),~(\ref{eq:8})) labels the {\it bps } running from left ($=\,1$) to right ($=\,100$). As the model depends on the relative positions between the pair mates, $GC$ following $GC$ cannot be distinguished from $GC$ following $CG$. {\it Closer to reality} descriptions should include 16 stacking interactions. The results for the sequence in Eq.~(\ref{eq:9}) are summarized in Fig.~\ref{fig:2} for a temperature window which features all the relevant denaturation effects. The numbers of path amplitudes exceeding $1{\AA}$, $1.5{\AA}$ and $2{\AA}$ respectively are plotted in Fig.~\ref{fig:2}(a) together with $N_\tau \times N_{eff}$ which ranges between $\sim 4.7 \cdot 10^6$ and $\sim 7 \cdot 10^6$ . The insets displays the path amplitudes normalized over $N_\tau \times N_{eff}$. All plots generally show a steady but not dramatic increase versus $T$ due to the dominance of strongly bounded $GC$ pairs. Some exceptions are however significant: at $T\sim 350K$, $N_\tau \times N_{eff}$ increases by over $2\cdot 10^5$ paths while two slightly less pronounced enhancements are found at $T\sim 310K$ ($1.6\cdot 10^5$ paths) and  $T\sim 375K$ ($10^5$ paths). The total number of paths contributing to $Z$ markedly increases when some groups of {\it bps } weaken their bonds signaling the interplay between  cooperativity and denaturation. These features are macroscopically seen in the plot of the specific heat (Fig.~\ref{fig:2}(c)) whereas the entropy (Fig.~\ref{fig:2}(b)) displays small irregularities at the same $T$ values and maintains an overall continuous behavior. The complementary microscopic explanation is provided by Fig.~\ref{fig:2}(d) which plots Eq.~(\ref{eq:8}) for two choices of the {\it threshold}, $\zeta=\,1{\AA}$ and $\zeta=\,1.5{\AA}$ respectively. In fact the fraction of mean stretchings exceeding $1{\AA}$, with respect to the double helix equilibrium configuration, shows somewhat appreciable jumps at about the same temperatures given above while the fraction larger than $1.5{\AA}$ becomes sizeable above $T \sim 325K$. Anyway $f$ never reaches the unity for such threshold values. The overall pictures emerging from Fig.~\ref{fig:2} is that of a continuous tendency towards denaturation essentially promoted by the $AT$ sites whose mean stretchings are generally larger than those for the GC pairs: this is made evident by the inset in Fig.~\ref{fig:2}(d) where Eq.~(\ref{eq:7}) is plotted for $i=\,5,15,25,45$. Also note that the $i=\,15, 25$ sites belong to a wider homogeneous AT region than the $i=\,5$ site hence the former display larger average elongations than the latter.

Now I take a AT-substrate in the right side of the fragment keeping the same sequence for the first 48 sites:

\begin{eqnarray}
GC &+& 6AT + GC + 22AT + 4GC + AT + 4GC  \nonumber
\\
&+& AT + 8GC + [49-100]AT
\label{eq:10}
\end{eqnarray}

\begin{figure}
\includegraphics[height=7.0cm,angle=0]{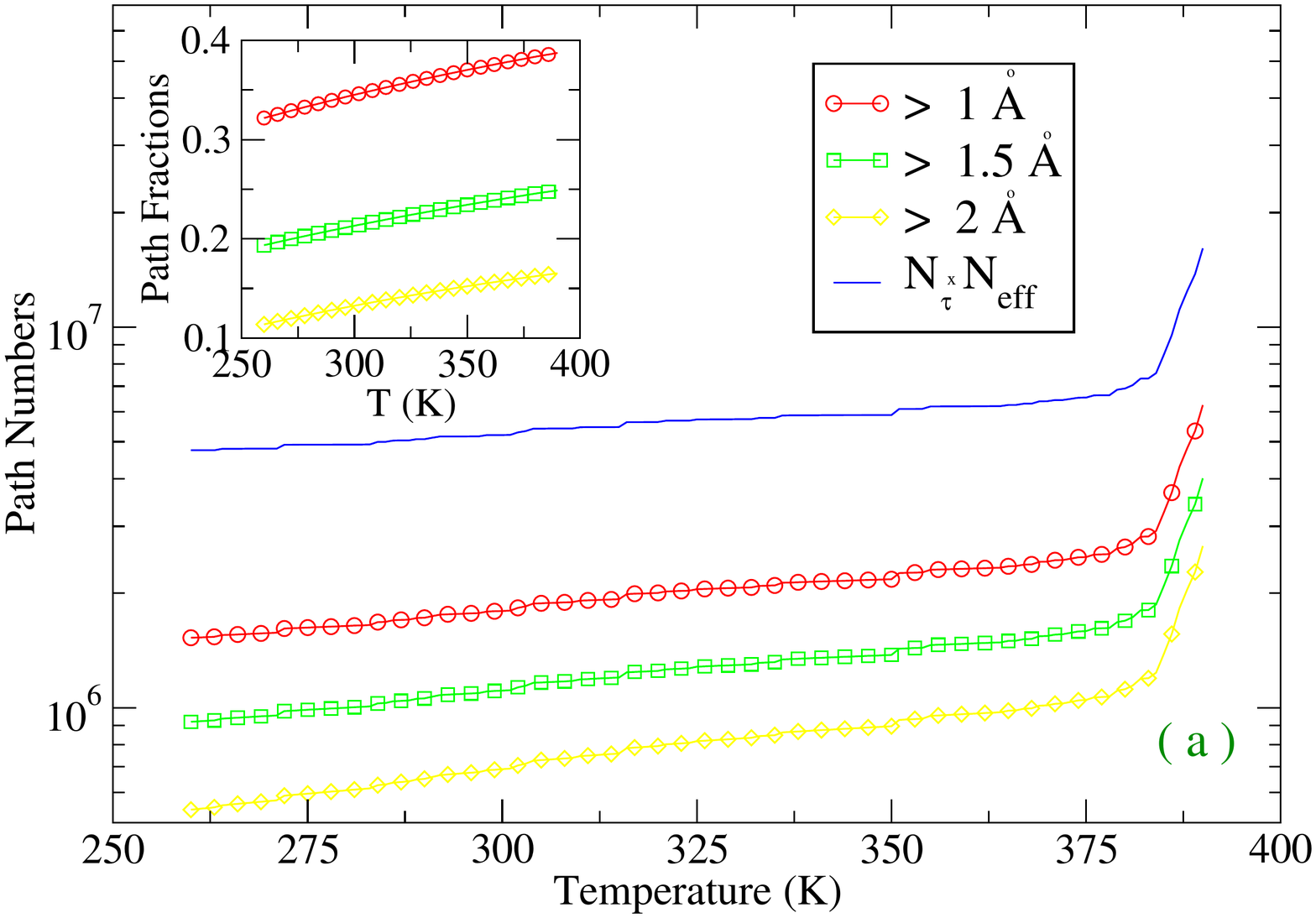}
\includegraphics[height=7.0cm,angle=0]{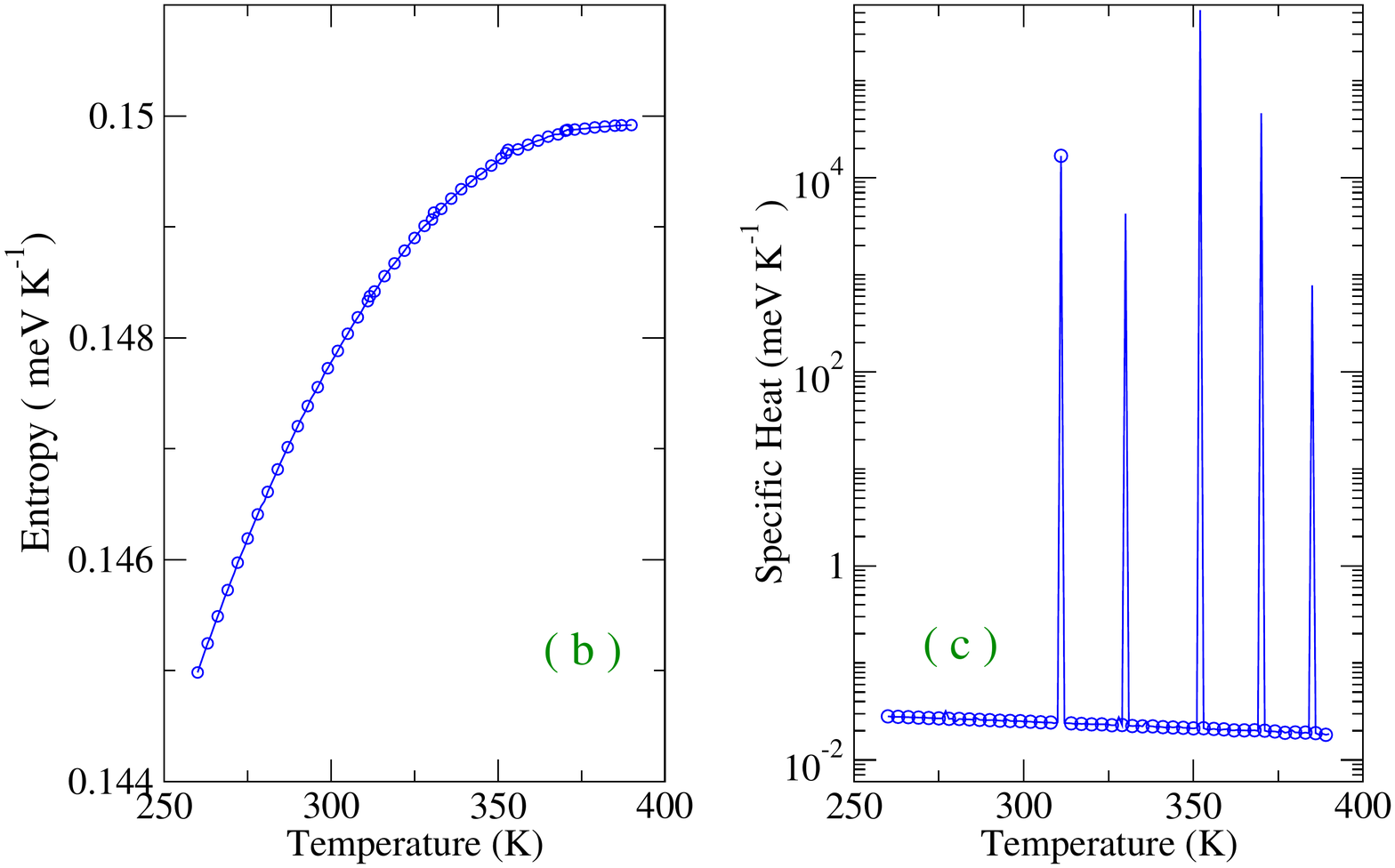}
\includegraphics[height=7.0cm,angle=0]{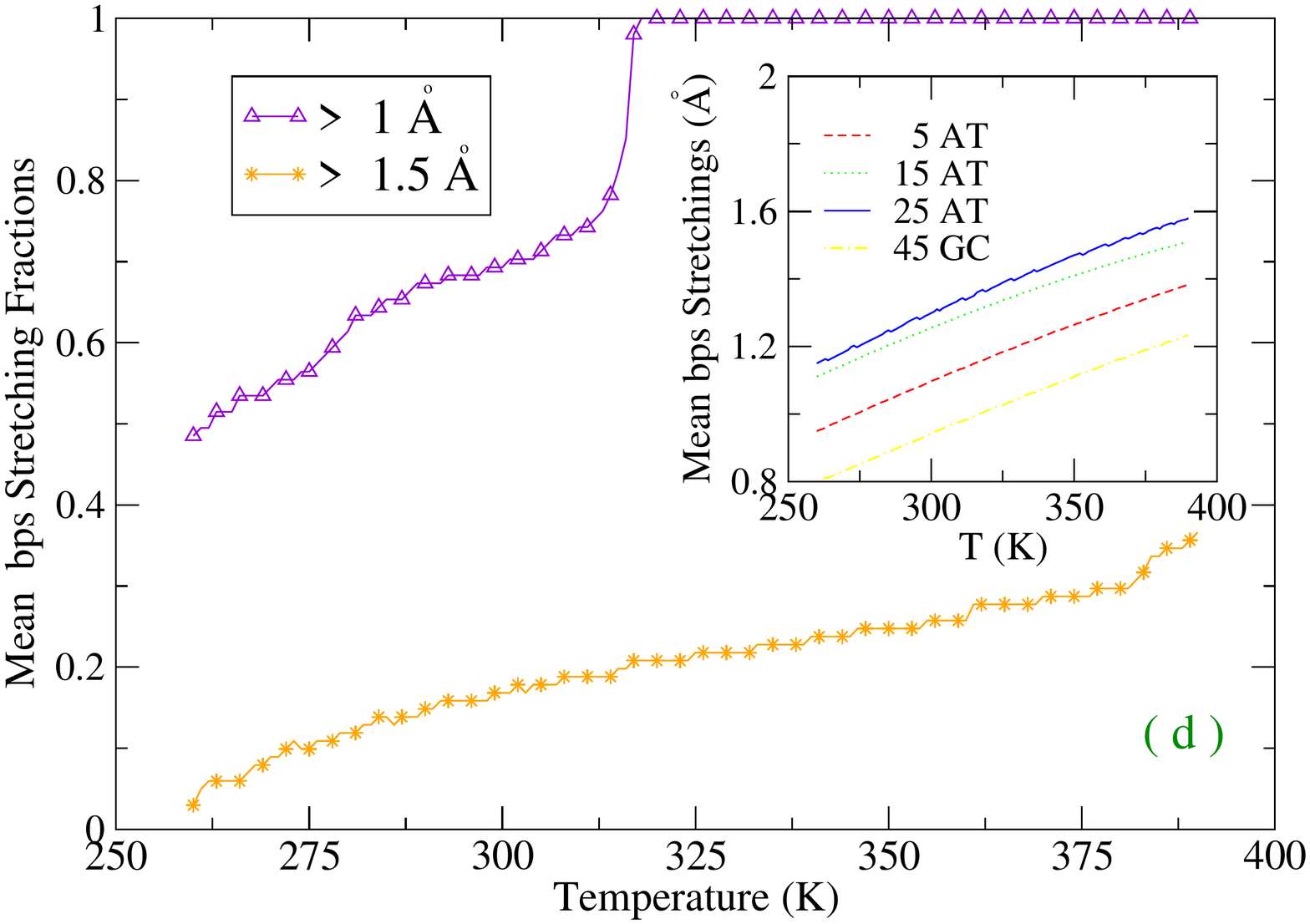}
\caption{\label{fig:3}(Color online) Sequence L48(AT30)+AT[49-100]. (a) Number of paths (for a single base pair stretching) larger than $1{\AA}$, $1.5{\AA}$ and $2{\AA}$;  Total Number of paths ($N_\tau \times N_{eff}$) contributing to the partition function. Inset: fractions of paths whose amplitude is larger than $1{\AA}$, $1.5{\AA}$ and $2{\AA}$ respectively.  (b) Entropy.  (c) Specific Heat. (d) Fractions of mean base pair stretchings calculated by Eq.~(\ref{eq:8}) for $\zeta=\, 1{\AA}$ and $\zeta=\,1.5{\AA}$ respectively. Inset: mean base pair stretchings at four specific sites. }
\end{figure}

The results for the fragment in Eq.~(\ref{eq:10}) are shown in Fig.~\ref{fig:3}. The portion of the path configuration space sampled by the computation is much larger than in the previous case with a strong increase at $T \sim 380K$ and $N_\tau \times N_{eff} \sim 18 \cdot 10^6$ at $T=\, 390K$. The path fractions exceeding $1{\AA}$, $1.5{\AA}$ and $2{\AA}$ respectively ( inset in Fig.~\ref{fig:3}(a) ) are similar to the previous case. However there is now a substantial increase of the absolute path numbers contributing to the denaturation with about three to six million path amplitudes broader than $1{\AA}$ in the upper temperature range. Consistently two more peaks appear in the specific heat plot beside the three ones already found in Fig.~\ref{fig:2}(c). Looking at  Fig.~\ref{fig:3}(d), we see that the fraction of mean base pair stretchings larger than $1{\AA}$ attains the unity at $T \sim 318K$ pretty close to the first peak encountered in the specific heat ($T \sim 311K$). As subsequent steps are found in the denaturation pattern at larger $T$ it may follow than $\zeta=\,1{\AA}$ underestimates the real threshold for the overall molecule denaturation. Or, the ensemble average procedure entering the definition of $f$ in  Eq.~(\ref{eq:8}) may not fully capture the occurrence of the molecule denaturation. While this issue deserves further work here we note that the mean path amplitudes at specific sites (inset in Fig.~\ref{fig:3}(d)) are significantly larger than those for the previous fragment (inset in Fig.~\ref{fig:2}(d)): this effect is due to the substrate made of weakly bound AT pairs. Even the $i=\,5$ AT site {\it feels} the change of the substrate (in spite of the distance along the fragment backbone) pointing to the importance of nonlocal cooperativity effects. Conversely the GC pairs at the first and third site may be viewed as the presence of two defects embedded in the AT-rich sequence to the left side. As the defects affect their surroundings  \cite{rapti} the $i=\,5$ site mean amplitude is somewhat smaller than that of other AT sites having homogeneous neighbors.

\begin{figure}
\includegraphics[height=7.0cm,angle=0]{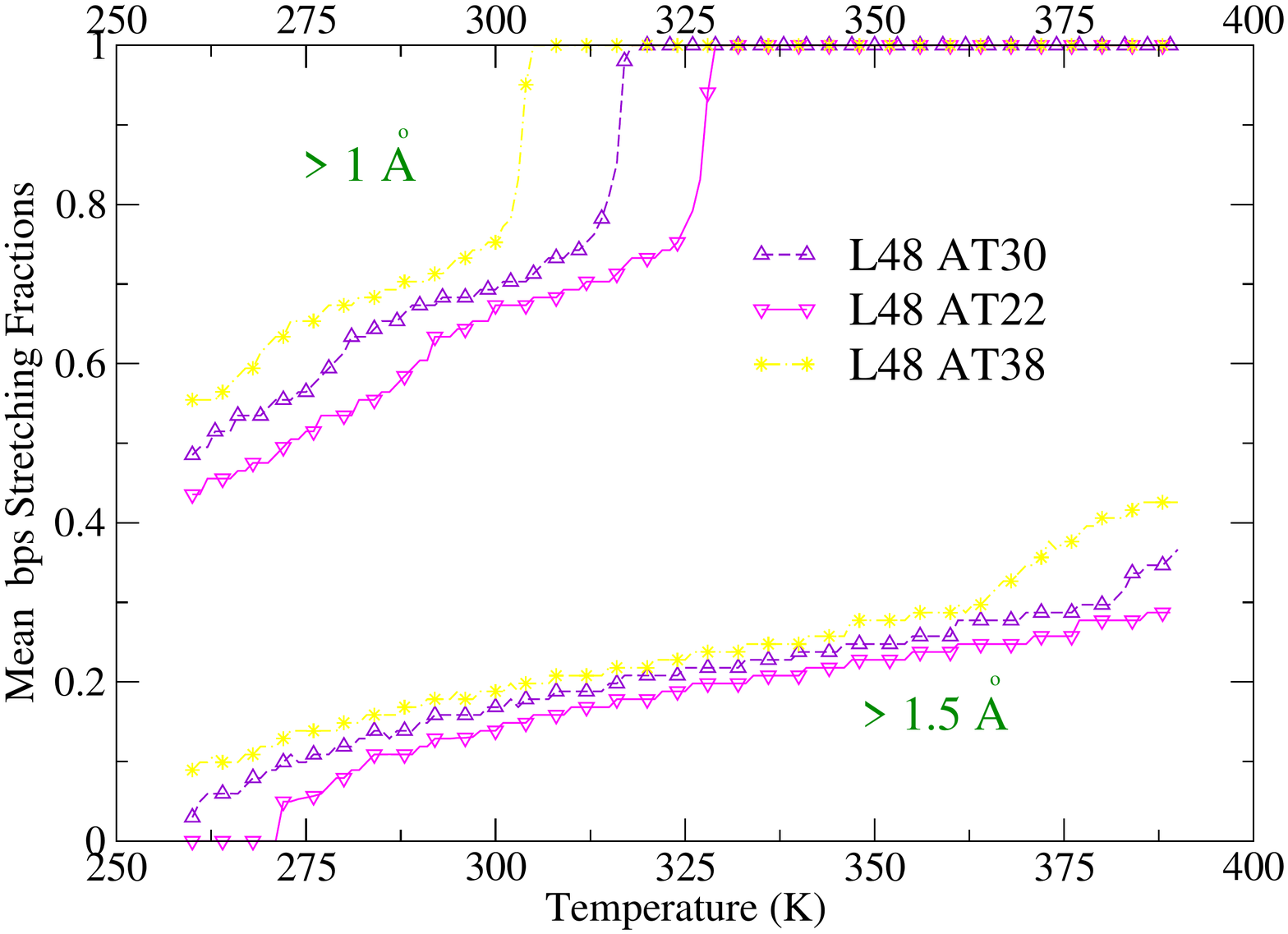}
\caption{\label{fig:4}(Color online) Fractions of mean base pair stretchings (Eq.~(\ref{eq:8})) larger than $1{\AA}$ and $1.5{\AA}$ respectively for three sequences : L48(AT30)+AT[49-100], L48(AT22)+AT[49-100] and L48(AT38)+AT[49-100]. }
\end{figure}

Eventually, the role of the AT {\it bps} is emphasized in Fig.~\ref{fig:4} where $f$ is computed for three cases: the AT sites content is increased/reduced by eight units (with respect to Fig.~\ref{fig:3}) in the first part of the sequence while the $[49-100]$ segment is kept fixed. The L48(AT22) sequence correspond to the L48AS sequence of Ref.\cite{zocchi1} which shows a two steps melting transition but a broad AT substrate is here attached to the sequence itself. Then, no direct comparison is possible. Again $f$ is computed by taking two threshold values as before. By adding (removing) eight AT sites, the temperature value such that $f$ attains the unity (for $\zeta=\,1{\AA}$) shifts downwards (upwards) by about $10K$. A significant increase in $f$ (for $\zeta=\,1.5{\AA}$) is also found at large $T$ for the AT-richest sequence.
Taking for good the functional form in Eq.~(\ref{eq:8}), a qualitative agreement is found with the melting profile calculated by Monte Carlo simulations of the PBD model \cite{ares} where the same definition for $f$ is assumed. A comparison between the L48(AT22) sequence in Fig.~\ref{fig:4} and the L48AS sequence in Ref.\cite{ares} suggests that a threshold $\zeta \sim \,1.1{\AA}$ permits to get $f=\,1$ at $T \sim 345K$ in both plots.

\section*{IV. Conclusion}

The temperature driven strands separation in heterogeneous DNA sequences has been studied by applying the path integral formalism to the nonlinear Peyrard-Bishop-Dauxois Hamiltonian model. Essentially the method consists in mapping the relative base pairs elongations onto the imaginary time scale set by the temperature. A time index $\tau_i$ labels each base pair which is thus described by all those paths, computed at $\tau_i$ in the path configuration space, which are compatible with the model potential and fulfill the macroscopic constraint given by the second law of thermodynamics.
The computational method requires that the entropy has to grow versus temperature but no \emph{ansatz} has been made regarding the shape of the entropy curves. The continuity found in the latter is consistent with the view that the strand separation is an overall smooth crossover similar in this respect to the case of homogeneous DNA.
The model has been applied to short fragments for which chain fluctuation effects are generally expected to broaden the transition region \cite{manghi}. In fact the molecule denaturation appears here as a multistep phenomenon, promoted by the AT-rich regions, whose long range effects may gradually extend over the whole fragment. The denaturation steps are signaled by a few significant enhancements in the number of paths which participate to the partition function although such enhancements are much less sharp than those previously found in homogeneous DNA. These findings are consistent with the fact that cooperativity is higher in homopolymers than in heteropolymers as, in the latter, different portions of the chain denaturate at different temperatures. The specific heat shows sharp peaks at about the same temperatures for which anomalies in the path numbers  plots occur. Beside a main transition peak  at $T\sim 350K$, our DNA sequences display some shoulder peaks whose frequency grows with a larger AT base pairs content. However some arbitrariness remains in the definition of the {\it threshold} for the occurrence of the overall molecule denaturation and much theoretical work remains to be done to unravel this issue.

The present conclusions regarding the smoothness of the denaturation are at variance with previous studies of the {PBD Hamiltonian} \cite{theo} suggesting that denaturation is a  first order thermodynamic transition microscopically driven by the backbone stiffness parameter both in homogeneous and heterogeneous sequences \cite{cule}. In fact the latter studies considered somewhat longer fragments than those I have taken but this should not be the source of the discrepancy regarding the character of the transition as the smooth crossover persists also by increasing the system size. Also some polymer network analysis  based on the {Poland-Scheraga model} for DNA \cite{peliti} point to a sharp denaturation which should be ascribed however to self-avoidance effects for the three dimensional molecule rather than to backbone stiffness. While the debate is open both inside the \emph{PBD Hamiltonian}- and the \emph{Poland-Scheraga model}- research fields, the path integral results here presented show that anharmonic stiffness alone should not change the character of the transition in heterogeneous DNA. Some improvements in modeling heterogeneous specific sequences are certainly expected by taking stiffness parameters which appropriately account for the stacking interactions along the molecule backbones.
This feature however is not expected to modify the nature of the crossover predicted by the path integral method. Instead, I feel that a main reason of divergence with respect to previous Hamiltonian studies lies in the fact that Eq.~(\ref{eq:3b}) incorporates all the path fluctuations around the ground state of the double strand structure. These fluctuations, included in the computational method, soften the effect of the entropic barrier associated to the stiffness and ultimately smoothen the crossover between the double strand configuration and a state with open domains.

Further, among the mesoscopic models capturing the essentials of the complex DNA interactions, the path integral method has the advantage to account for a remarkable number of molecule configurations in a short computational time. Nonetheless some limitations regarding both model and method should be here recognized with the purpose to be lifted in next investigations. First, the path integral in Eq.~(\ref{eq:3b}) describes a one dimensional system: extensions to higher dimensionality may permit to fully include self-avoiding paths in the computation. Second, the space-time mapping technique in Eq.~(\ref{eq:3}) may be modified by removing the correspondence between \emph{one} base pair and \emph{one} point $\tau_i$ along the imaginary time axis. By freeing the time from such constraint, each base pair stretching would maintain the full time dependence and \emph{one} point in the path configuration space would correspond to \emph{one} molecule whose different configurations could then be obtained by tuning the time. Accordingly the configuration space spanned by the computation would describe an ensemble of molecules each of them existing in an ensemble of different states. In this way the length of the molecules in the ensemble would become a free parameter thus allowing us to examine the denaturation process both for long DNA chains and  fragments with only a few tens of base pairs. Analysis of long sequences may permit to check the role of the path fluctuations approaching the thermodynamic limit.  On the other hand, short fragments are particularly interesting also in view of the fact that experiments capable to detect intermediate states in the melting transition are becoming available.

\section*{Acknowledgements}

I wish to thank Dr. G. Costantini for prompt and skillful collaboration.


\begin{references}
\bibitem{note}
Due to the on site potential with unbounded transverse stretching $y_n$, a thermodynamic phase transition is possible also in the harmonic ($\rho=\,0$) Peyrard-Bishop Hamiltonian \cite{pey1}. Thus there is no contradiction with general theorems regarding the impossibility of phase transitions in one dimensional systems \cite{hove}.
\end{references}
\end{document}